%% file: main.tex
\documentclass{article}
\PassOptionsToPackage{numbers, compress}{natbib}
\usepackage[preprint]{neurips_2025}

\usepackage[utf8]{inputenc}
\usepackage[T1]{fontenc}
\usepackage{fix-cm}
\usepackage{amsmath,amssymb,amsfonts}
\usepackage{graphicx}
\usepackage{booktabs}
\usepackage{nicefrac}
\usepackage{microtype}
\usepackage{xcolor}
\definecolor{linkColor}{rgb}{0.2,0.4,0.6}
\usepackage[colorlinks=true,linkcolor=linkColor,citecolor=linkColor,filecolor=linkColor,urlcolor=linkColor]{hyperref}
\usepackage{url}
\usepackage{caption}
\usepackage{subcaption}
\usepackage{array}
\usepackage{multirow}
\usepackage{enumitem}
\usepackage{float}
\usepackage{placeins}
\usepackage{cleveref}
\usepackage{tcolorbox}

\definecolor{abstractbg}{gray}{0.95}
\makeatletter
\renewenvironment{abstract}{%
  \vskip 0.1in
  \begin{tcolorbox}[
    colback=abstractbg, colframe=abstractbg,
    arc=3mm, boxrule=0pt,
    left=6mm, right=6mm, top=4mm, bottom=4mm
  ]
}{%
  \end{tcolorbox}
}
\makeatother

\newcommand{\vibeasr}{\textsc{VibeVoice-ASR-BitNet}}
\newcommand{\vibebase}{\textsc{VibeVoice-ASR}}

\newcommand{\ocmark}{\raisebox{-0.1em}{\includegraphics[height=0.9em]{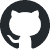}}}
\newcommand{\hfmark}{\raisebox{-0.1em}{\includegraphics[height=0.9em]{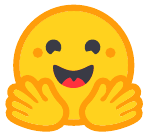}}}

\title{\vibeasr{} Technical Report}

\author{
\textnormal{Songchen Xu$^{*}$ \quad Ting Song \quad Shaohan Huang \quad Zhiliang Peng \quad Yan Xia \quad Yujie Tu} \\
\textnormal{Xin Huang \quad Xun Wu \quad Wenhui Wang \quad Yaoyao Chang \quad Jianwei Yu \quad Li Dong \quad Furu Wei$^{\diamond}$} \\[4pt]
\url{https://aka.ms/GeneralAI}
}

\begin{document}

\maketitle

\begin{abstract}
\input{sections/abstract}
\end{abstract}

\vspace{-0.3em}
\begin{center}
\small
\ocmark{} Code: \href{https://github.com/microsoft/VibeASR.cpp}{\texttt{github.com/microsoft/VibeASR.cpp}} \\[2pt]
\hfmark{} Models: \href{https://huggingface.co/microsoft/VibeVoice-ASR-BitNet}{\texttt{huggingface.co/microsoft/VibeVoice-ASR-BitNet}}
\end{center}
\vspace{-0.5em}

\renewcommand{\thefootnote}{}
\footnotetext{\hspace{-1.8em}$^{*}$Main contribution. $\diamond$\,Corresponding author: \href{mailto:fuwei@microsoft.com}{fuwei@microsoft.com}. Z. Peng, T. Song, Y. Xia, S. Huang, X. Wu, W. Wang, Y. Chang, J. Yu, L. Dong and F. Wei are with Microsoft Research. S. Xu is with Shanghai Jiao Tong University. X. Huang is with Fudan University. Y. Tu is with University of Chinese Academy of Sciences.}
\renewcommand{\thefootnote}{\arabic{footnote}}

\begin{figure}[h!]
    \centering
    \includegraphics[width=0.92\linewidth]{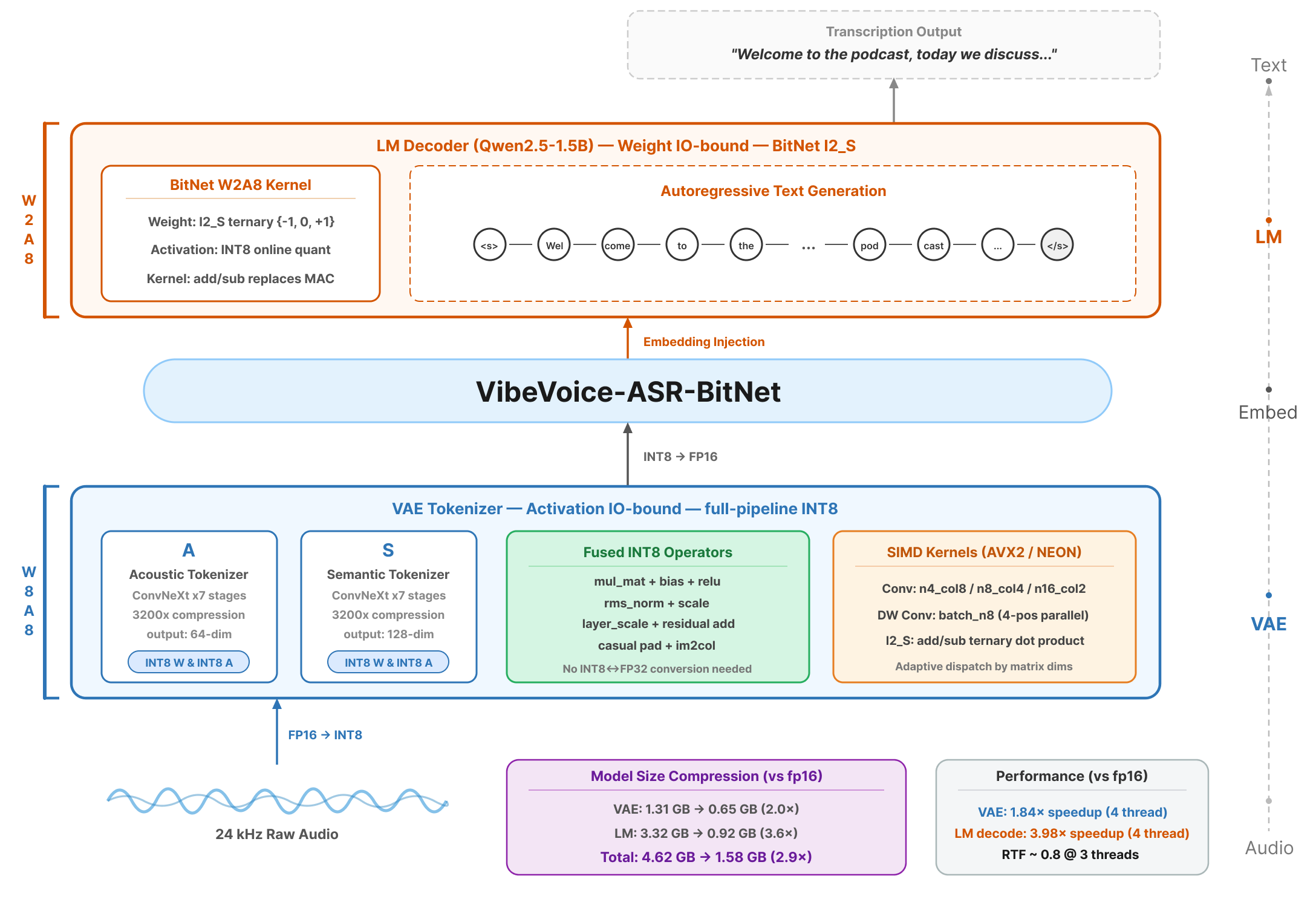}
    \vspace{-0.5em}
    \caption{Overview of the \vibeasr{} system: heterogeneous quantization enables real-time CPU inference with $2.9\times$ model compression relative to FP16.}
    \label{fig:overview}
\end{figure}

\newpage
\input{sections/introduction}
\input{sections/method}
\input{sections/result}
\input{sections/conclusion}

\newpage
\bibliographystyle{plainnat}
\bibliography{references}

\end{document}

%% file: sections/abstract.tex
We present VibeVoice-ASR-BitNet, a compressed variant of VibeVoice-ASR optimized for real-time inference on edge CPUs. We apply heterogeneous quantization tailored to the computational characteristics of each stage: the VAE acoustic tokenizer uses full-pipeline INT8 quantization (I8\_S) with kernel fusion and SIMD optimization, while the autoregressive language model adopts BitNet-style ternary weights (I2\_S). To preserve accuracy under aggressive compression, we employ a progressive quantization-aware training strategy. For inference, we implement custom SIMD kernels and fused operators within the ggml framework targeting both ARM and x86 platforms, achieving real-time recognition (RTF $< 1$) on low-thread-count CPUs. VibeVoice-ASR-BitNet is $1.6$--$2.3\times$ faster than Whisper.cpp at comparable model sizes ($\sim$1.6\,GB), with only modest accuracy degradation compared to the FP16 baseline.

%% file: sections/introduction.tex
\section{Introduction}
\label{sec:introduction}

Automatic Speech Recognition (ASR) has advanced rapidly in recent years, with two dominant paradigms emerging. Traditional lightweight approaches---such as transducer models \citep{graves2012transducer}, CTC-based systems \citep{graves2006ctc}, and compact encoder-decoder architectures like Conformer \citep{gulati2020conformer} and Zipformer \citep{yao2024zipformer}---offer fast inference suitable for streaming scenarios, but typically suffer from limited accuracy on diverse domains and lack robust multilingual capabilities. In contrast, Large Language Model (LLM)-based ASR systems \citep{radford2023whisper, peng2025vibevoiceasr, zhang2024moonshine, pratap2024scaling} achieve state-of-the-art accuracy by leveraging the powerful sequence modeling capacity of billion-parameter transformers, but at the cost of significantly higher computational demands.

LLM-based ASR models are predominantly deployed on GPUs in cloud environments. Systems such as Whisper \citep{radford2023whisper}, Qwen-Audio \citep{chu2023qwenaudio}, SeamlessM4T \citep{barrault2023seamlessm4t}, and VibeVoice-ASR \citep{peng2025vibevoiceasr} rely on server-grade hardware to meet latency requirements. However, cloud-based inference introduces privacy concerns---sensitive audio data must leave the device---and network-dependent latency that is unacceptable for real-time applications. While CPU-based inference engines such as Whisper.cpp \citep{whispercpp2023} and llama.cpp \citep{llamacpp2023} have made local deployment feasible, they typically fail to achieve real-time performance (RTF $< 1$) for large models on resource-constrained devices, or require high thread counts that saturate the CPU budget of edge hardware.

In this work, we present VibeVoice-ASR-BitNet (Figure~\ref{fig:overview}), which addresses these challenges through heterogeneous quantization co-designed with the computational profile of each model component. The VAE tokenizer, encompassing both acoustic and semantic encoders, is IO-bound due to its convolutional architecture and is quantized to full-pipeline I8\_S to reduce memory bandwidth by $2\times$ relative to FP16. The autoregressive language model adopts BitNet-style ternary weights I2\_S \citep{ma2024bitnet}, compressing weights by $8\times$ relative to FP16 and leveraging INT8 multiply-add kernels to reduce memory bandwidth pressure during both prefill and decode stages. Combined with custom SIMD kernels and operator fusion in the ggml framework \citep{ggml2023}, the total model size is compressed from 4.62\,GB to 1.58\,GB, a $2.9\times$ reduction, and the system achieves real-time inference (RTF $< 1$) on commodity CPUs without requiring high thread counts, while maintaining competitive accuracy with only modest degradation relative to the FP16 baseline.

%% file: sections/method.tex
\section{Method}
\label{sec:method}

\subsection{Model Architecture Overview}
\label{sec:model_arch}

\begin{figure}[ht]
    \centering
    \includegraphics[width=\linewidth]{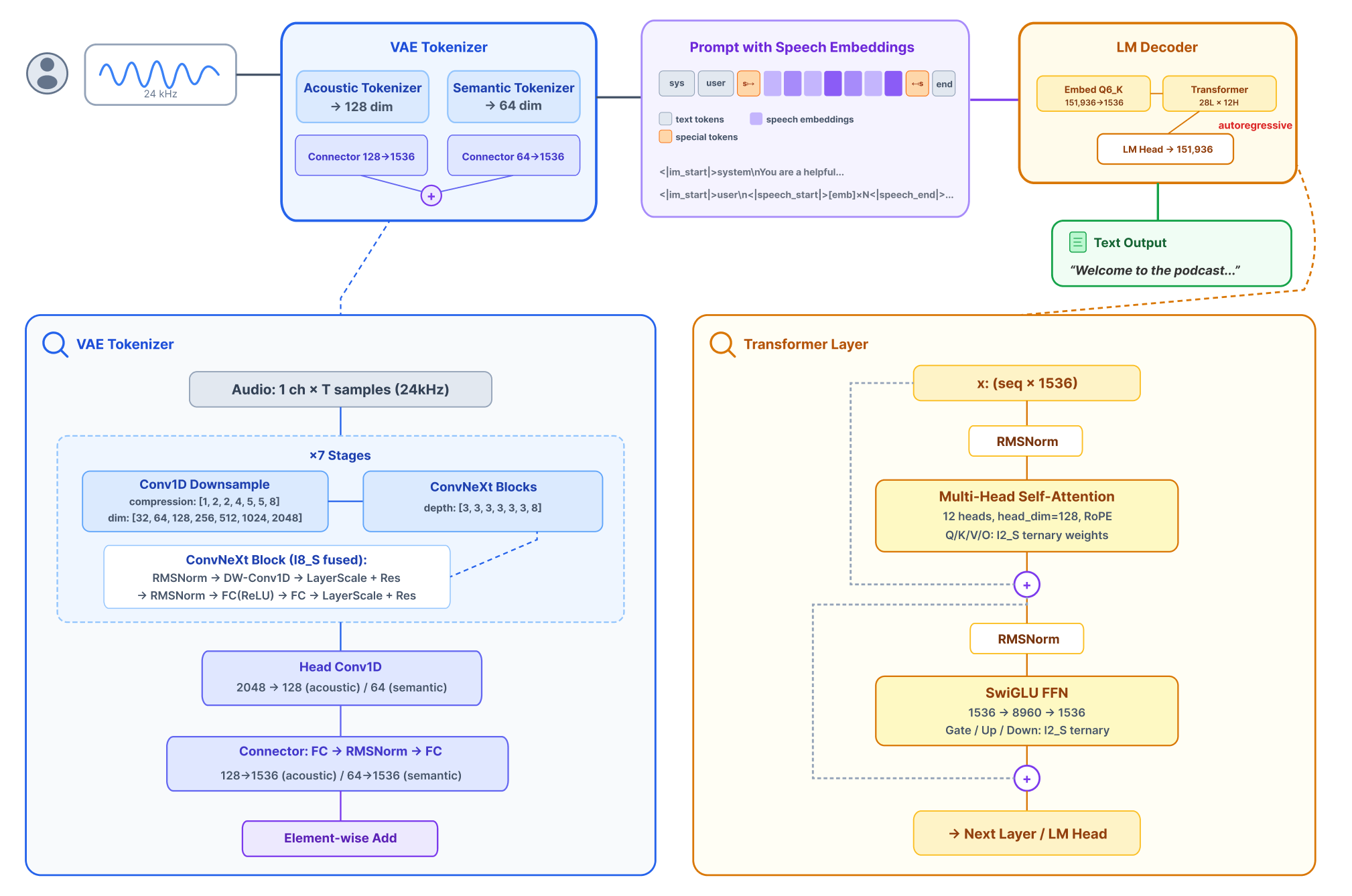}
    \caption{Model architecture of VibeVoice-ASR-BitNet. Left: VAE tokenizer, 7-stage ConvNeXt with I8\_S fused operators. Right: LM decoder, 28-layer Transformer with I2\_S ternary weights.}
    \label{fig:arch}
\end{figure}

VibeVoice-ASR-BitNet inherits the architecture of VibeVoice-ASR \citep{peng2025vibevoiceasr}: a VAE tokenizer---acoustic encoder + semantic encoder, both 7-stage ConvNeXt \citep{liu2022convnext} backbones with $3200\times$ temporal downsampling from 24\,kHz to 7.5\,Hz---followed by an autoregressive language model decoder. To reduce memory footprint for edge CPU deployment, we replace the original Qwen2.5-7B \citep{qwen2024qwen25} decoder with Qwen2.5-1.5B. The full architecture is shown in Figure~\ref{fig:arch}. We refer readers to \citet{peng2025vibevoiceasr} for details of the base model.

\subsection{Heterogeneous Quantization}
\label{sec:quantization}

The two stages of VibeVoice-ASR exhibit fundamentally different computational profiles, motivating a heterogeneous quantization strategy rather than a uniform precision reduction.

\subsubsection{I8\_S: Full-Pipeline INT8 for VAE Tokenizer}
\label{sec:i8s}

\paragraph{Memory Traffic Analysis.}
Each ConvNeXt block consists of a depthwise mixer convolution (implemented via \texttt{im2col} + matrix multiply) and a two-layer FFN ($D \to 4D \to D$). For channel dimension $D$, sequence length $L$, and kernel size $k=8$, the per-block weight parameter count and activation data volume are:
\begin{align}
    W &= \underbrace{Dk}_{\text{mixer}} + \underbrace{8D^2}_{\text{FFN}} \label{eq:weight} \\[4pt]
    A &= \underbrace{2DkL}_{\text{im2col}} + \underbrace{12DL}_{\text{FFN I/O}} \label{eq:activation}
\end{align}
Specifically, $W$ sums the depthwise convolution kernel ($D \times k$ parameters) and the two FFN matrices ($D \times 4D + 4D \times D = 8D^2$). $A$ sums the \texttt{im2col} read/write traffic ($D \times k \times L$ elements read, plus $D \times k \times L$ elements written to the intermediate buffer) and all FFN activation I/O ($4DL$ per linear layer $\times$ 3 passes: input, intermediate, output). Table~\ref{tab:traffic} shows the per-stage breakdown for a 20\,s audio input (480k samples at 24\,kHz).

\begin{table}[ht]
\centering
\small
\begin{tabular}{crrrrrr}
\toprule
Stage & $D$ & $L$ & Depth & Weights (M) & Activations (M) & Act\,/\,W \\
\midrule
0 & 32 & 480,000 & 3 & 0.03 & 1,290 & 50,909 \\
1 & 64 & 240,000 & 3 & 0.10 & 1,290 & 12,923 \\
2 & 128 & 120,000 & 3 & 0.40 & 1,290 & 3,256 \\
3 & 256 & 30,000 & 3 & 1.6 & 645 & 409 \\
4 & 512 & 6,000 & 3 & 6.3 & 258 & 41 \\
5 & 1,024 & 1,200 & 3 & 25.2 & 103 & 4 \\
6 & 2,048 & 150 & 8 & 268.6 & 69 & 0.3 \\
\midrule
\multicolumn{4}{c}{\textbf{Total}} & \textbf{302} & \textbf{4,946} & \textbf{16.4} \\
\bottomrule
\end{tabular}
\caption{Per-stage weight parameters and activation data volume of the VAE tokenizer (20\,s audio, $k=8$).}
\label{tab:traffic}
\end{table}

The Act/W ratio reveals that activation data volume exceeds weight parameters by $16.4\times$ in aggregate, and by up to $50{,}000\times$ for early stages. This means weight-only quantization (e.g., W8A32) leaves the dominant memory traffic path uncompressed, yielding negligible speedup. Furthermore, activation data must be kept in INT8 throughout the entire forward pass---if activations were stored in higher precision and quantized on-the-fly at each layer boundary, the conversion overhead on billions of elements would be prohibitive. I8\_S therefore enforces a uniform INT8 datapath across the entire forward pass---weights, activations, intermediate buffers, and all operator boundaries operate in INT8, eliminating all precision format conversions.

\paragraph{Why INT8 and Not Lower.}
Sub-byte formats (e.g., INT4) require pack/unpack operations that partially offset IO savings, lack native SIMD instruction support on most CPU architectures, and cause rapid accuracy degradation in the ConvNeXt architecture. INT8 provides native hardware support (AVX2/NEON) with negligible accuracy loss.

\subsubsection{I2\_S: BitNet Ternary Weights for Language Model}
\label{sec:i2s}

The LM decoder is weight-dominated during autoregressive decoding---each token reads the full 1.5B weight matrix but only a single activation vector. We adopt the BitNet approach \citep{ma2024bitnet}: weights are quantized to ternary $\{-1, 0, +1\}$ stored in 2 bits, and activations are quantized to INT8 online per token. This reduces weight memory traffic by $8\times$ vs.\ FP16. The embedding and LM head layers are quantized to Q6\_K (6-bit) rather than ternary, since these layers map between the full vocabulary space and hidden states, where ternary precision causes unacceptable degradation. Following the multiply-add computation method in BitNet, our I2\_S kernel unpacks ternary weights to INT8 values and computes via the same \texttt{maddubs} pipeline as I8\_S, unifying the kernel design for both quantization schemes while retaining the memory bandwidth advantage of 2-bit weight packing.

\subsection{Progressive Quantization-Aware Training}
\label{sec:qat}

\subsubsection{VAE Tokenizer}

The VAE tokenizer undergoes two modifications before QAT. First, all GELU activations are replaced with ReLU, since GELU is difficult to implement efficiently under INT8 arithmetic. The model is briefly finetuned after this substitution to recover accuracy.

\begin{figure}[ht]
\centering
\includegraphics[width=0.85\linewidth]{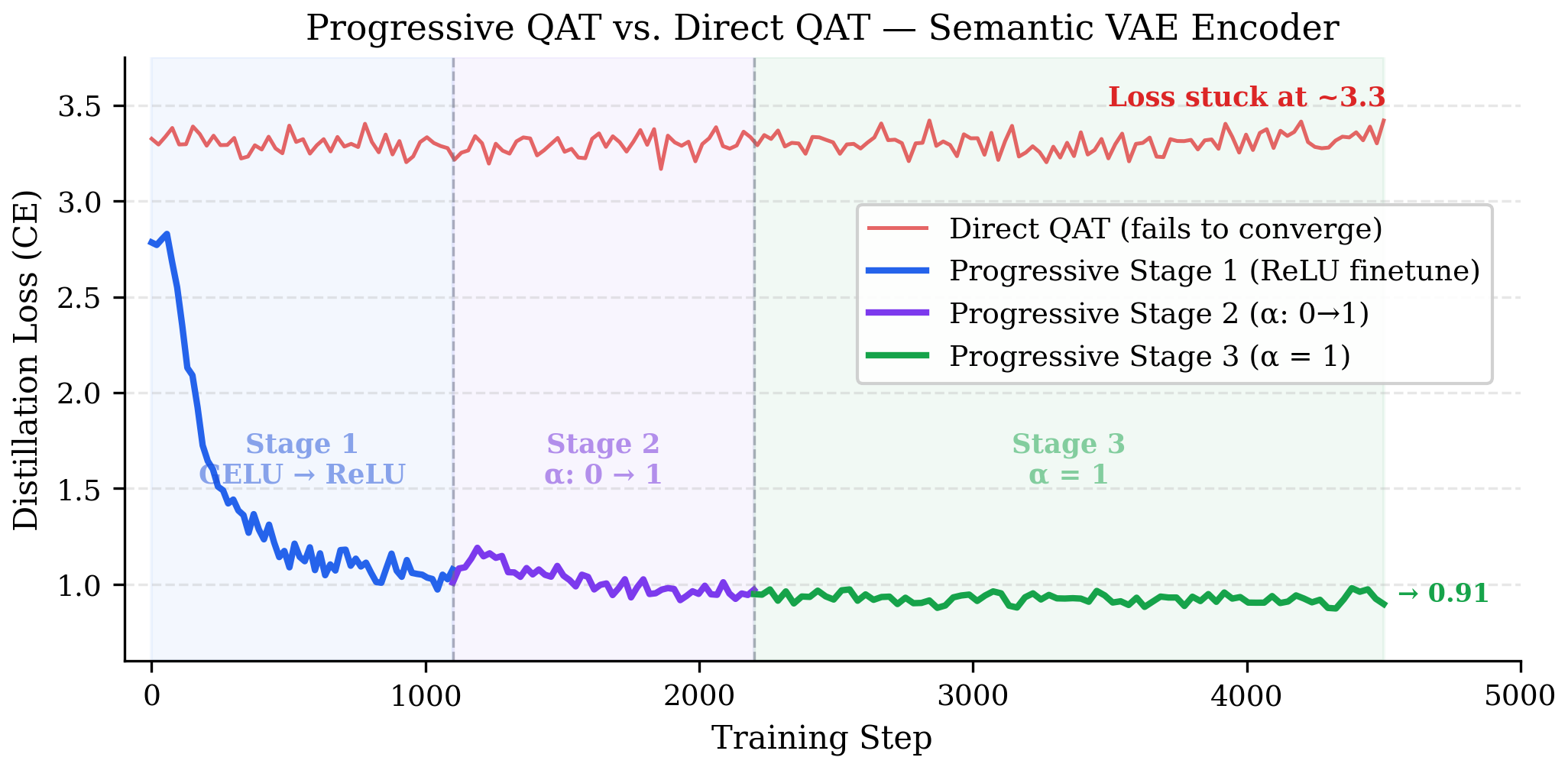}
\caption{Training loss comparison: direct QAT fails to converge (loss $\approx 3.3$), while progressive QAT converges to $\approx 0.91$ in three stages (GELU$\to$ReLU finetune, $\alpha$: 0$\to$1 blending, $\alpha = 1$ full quantization).}
\label{fig:training_curves}
\end{figure}

Second, fake-quantize nodes are inserted at all weight and activation boundaries. To avoid the training instability caused by direct quantization, we adopt a progressive schedule controlled by a blending parameter $\alpha \in [0, 1]$:
\begin{equation}
    \hat{x} = (1 - \alpha) \cdot x + \alpha \cdot \text{FakeQuant}(x)
    \label{eq:progressive}
\end{equation}
where $\alpha$ is linearly increased from 0 to 1 over the course of training. At $\alpha = 0$ the model operates in full precision; at $\alpha = 1$ it is fully quantized. Once $\alpha$ reaches 1, training continues at full quantization to further stabilize the quantized weights. This three-stage progressive transition ensures stable loss descent---in our experiments, direct QAT ($\alpha = 1$ from the start) fails to converge, with loss oscillating around 3.3 indefinitely. Figure~\ref{fig:training_curves} compares the two strategies.

\subsubsection{Language Model}

To reduce memory footprint for edge CPU deployment, we replace the original Qwen2.5-7B decoder in VibeVoice-ASR with Qwen2.5-1.5B. Given the smaller model capacity and the fact that edge CPU real-time ASR typically operates on chunked audio rather than processing entire long recordings in a single pass, we accordingly limit training data to segments under 4 minutes, unlike the original VibeVoice-ASR which supports 60-minute long-form audio. Despite the $4.7\times$ parameter reduction and shorter training sequences, experiments show only modest accuracy degradation on ASR tasks (typically 1--4\% absolute WER increase; see Section~\ref{sec:accuracy}), confirming that a 1.5B-class LM with shorter context is sufficient for edge speech recognition.

After the INT8 VAE tokenizer is trained, it is frozen and integrated into the full ASR pipeline. The LM decoder is then trained with BitNet ternary weights following the two-stage procedure of VibeVoice-ASR \citep{peng2025vibevoiceasr}: large-scale pretraining on speech-text pairs followed by supervised finetuning on target domains.

\subsection{Inference Optimization}
\label{sec:inference}

\subsubsection{Operator Fusion}

To reduce intermediate memory traffic, we fuse frequently co-occurring operators into single kernels. The fused operators for the VAE tokenizer include:

\begin{itemize}[nosep,leftmargin=1.5em]
    \item \textbf{im2col\_asym}: fused asymmetric padding + im2col expansion
    \item \textbf{mul\_mat\_add\_relu}: fused matrix multiply + bias + ReLU activation
    \item \textbf{add\_scaled}: fused residual addition with layer-scale multiplication
    \item \textbf{rms\_norm\_scaled}: fused RMS normalization with scale parameter
\end{itemize}

Each fused kernel eliminates one or more intermediate tensor materializations, avoiding redundant memory round-trips. For the VAE tokenizer where activation IO dominates (Section~\ref{sec:i8s}), this fusion is critical---it reduces the effective activation traffic by eliminating temporary buffers between adjacent operations.

\subsubsection{Custom SIMD Kernels}

\begin{figure}[ht]
    \centering
    \includegraphics[width=\linewidth]{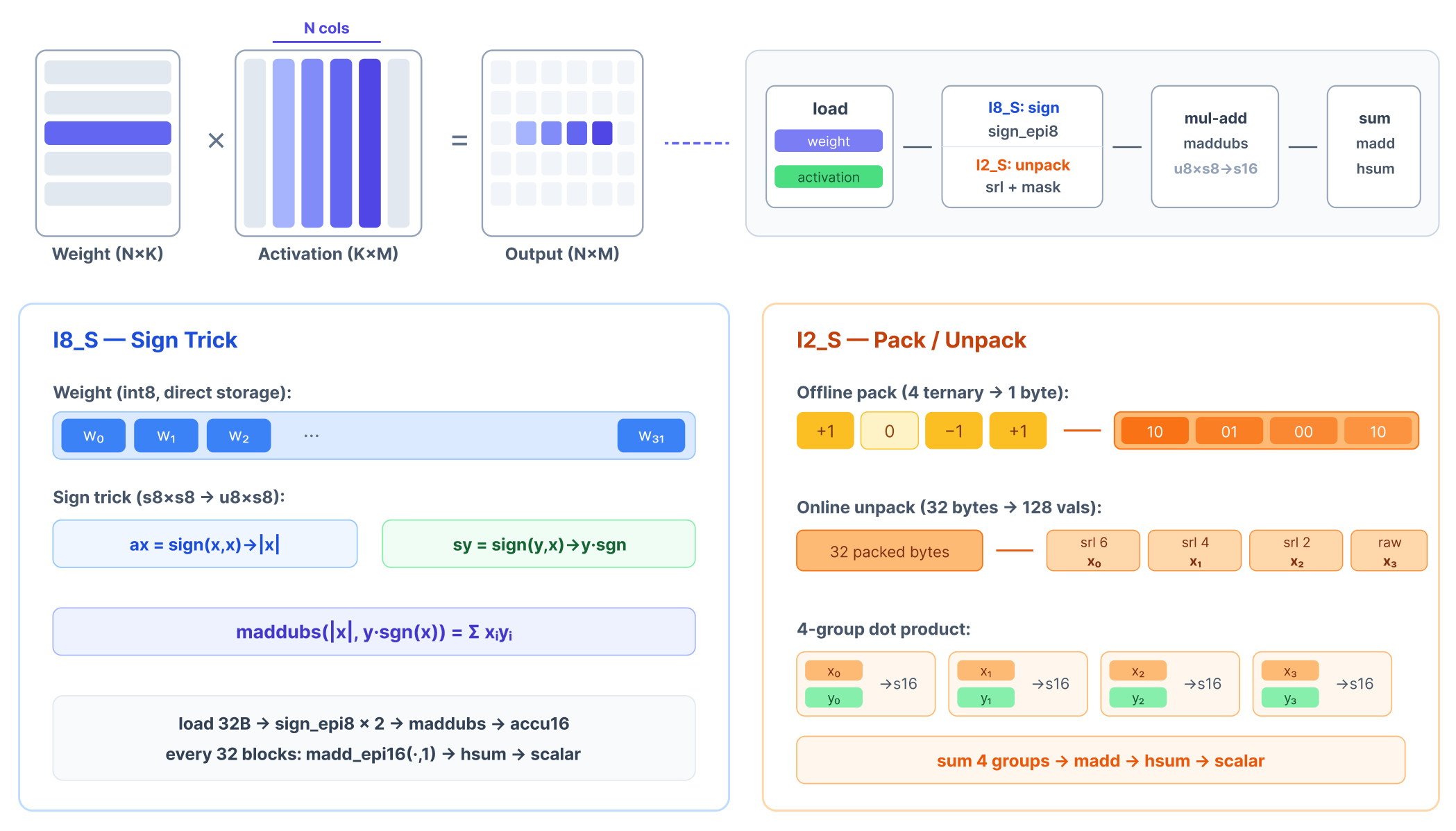}
    \caption{I8\_S and I2\_S GEMM kernel: N$\times$1 parallel \texttt{vec\_dot} with SIMD multiply-add instructions. Both share the same \texttt{maddubs} $\to$ \texttt{madd} $\to$ \texttt{hsum} accumulation pipeline; I8\_S uses the sign trick for signed multiplication, while I2\_S adds an online unpack stage to convert 2-bit packed weights to INT8 before entering the same pipeline.}
    \label{fig:gemm}
\end{figure}

\paragraph{I8\_S / I2\_S GEMM Kernel.}
The I8\_S kernel processes data in blocks of $Q_K = 32$ elements, using \texttt{\_mm256\_maddubs\_epi16} on AVX2 or NEON \texttt{smull}/\texttt{sadalp} to perform 32 INT8 multiply-add operations per instruction. The I2\_S kernel unpacks 2-bit ternary values from packed bytes into INT8 format, then feeds them through the same \texttt{maddubs} pipeline, with $Q_K = 128$ on x86 or $Q_K = 64$ on ARM. Figure~\ref{fig:gemm} illustrates the N$\times$1 parallel \texttt{vec\_dot} dataflow for both kernels. Both kernels support multiple tiling strategies---\texttt{1$\times$N}, \texttt{N$\times$1}, and batch variants---selected at runtime based on matrix dimensions to maximize register utilization.

\paragraph{Other Kernels.}
Additional SIMD-optimized kernels include INT8 RMS normalization and layer-scale multiplication, which complement the GEMM kernels to form a complete INT8 computation pipeline.

%% file: sections/result.tex
\section{Experiments}
\label{sec:result}

\subsection{Experimental Setup}
\label{sec:setup}

\paragraph{Platform.}
All experiments are conducted on a server equipped with an AMD EPYC 7V13 processor (24 cores, 3.1\,GHz base clock, no SMT), 216\,GB DDR4 memory, and 96\,MB L3 cache. The CPU supports AVX2 and FMA instruction sets.

\paragraph{Datasets.}
We evaluate on two categories of benchmarks: (1) MLC (Multilingual LibriSpeech Corpus) covering 6 languages---English, French, Italian, Korean, Portuguese, and Vietnamese; (2) standard benchmarks including AISHELL4, AMI (ihm/sdm), AliMeeting, Fleurs (English/Chinese), LibriSpeech (clean/other), and VoxPopuli. These cover read speech, meeting/conversational speech, and multilingual scenarios.

\paragraph{Metrics.}
We report Character Error Rate (CER) and Word Error Rate (WER) for accuracy evaluation. For inference speed, we use Real-Time Factor (RTF = inference time / audio duration); RTF $< 1$ indicates real-time capability.

\subsection{Inference Performance}
\label{sec:performance}

Table~\ref{tab:model_size} summarizes the model size before and after heterogeneous quantization. The total model footprint is reduced from 4.62\,GB (FP16) to 1.58\,GB, a $2.9\times$ compression that enables deployment on memory-constrained edge devices.

\begin{table}[ht]
\centering
\small
\begin{tabular}{lccc}
\toprule
\textbf{Component} & \textbf{FP16} & \textbf{Quantized} & \textbf{Compression} \\
\midrule
VAE Tokenizer & 1.31\,GB & 0.65\,GB (I8\_S) & 2.0$\times$ \\
LM Decoder & 3.32\,GB & 0.92\,GB (I2\_S + Q6\_K embed) & 3.6$\times$ \\
\midrule
Total & 4.62\,GB & 1.58\,GB & 2.9$\times$ \\
\bottomrule
\end{tabular}
\caption{Model size before and after quantization.}
\label{tab:model_size}
\end{table}

Table~\ref{tab:rtf_e2e} reports the end-to-end RTF across different audio durations and thread counts. The system achieves real-time inference (RTF $< 1$) at low thread counts across all tested durations (5--40\,s). RTF decreases slightly as audio length increases, because the fixed overhead of pipeline initialization is amortized over longer inputs.

\begin{table}[ht]
\centering
\small
\begin{tabular}{ccccccc}
\toprule
Audio Duration & 1 Thread & 2 Threads & 3 Threads & 4 Threads & 6 Threads & 8 Threads \\
\midrule
5\,s  & 2.11 & 1.22 & \textbf{0.89} & \textbf{0.76} & \textbf{0.60} & \textbf{0.54} \\
10\,s & 2.05 & 1.13 & \textbf{0.82} & \textbf{0.69} & \textbf{0.53} & \textbf{0.47} \\
20\,s & 1.98 & 1.08 & \textbf{0.77} & \textbf{0.63} & \textbf{0.49} & \textbf{0.42} \\
40\,s & 1.96  & 1.05 & \textbf{0.76} & \textbf{0.61} & \textbf{0.47} & \textbf{0.41} \\
\bottomrule
\end{tabular}
\caption{End-to-end Real-Time Factor under different audio durations and thread counts. Bold indicates real-time, RTF $< 1$.}
\label{tab:rtf_e2e}
\end{table}

\begin{figure}[ht]
\centering
\includegraphics[width=0.75\linewidth]{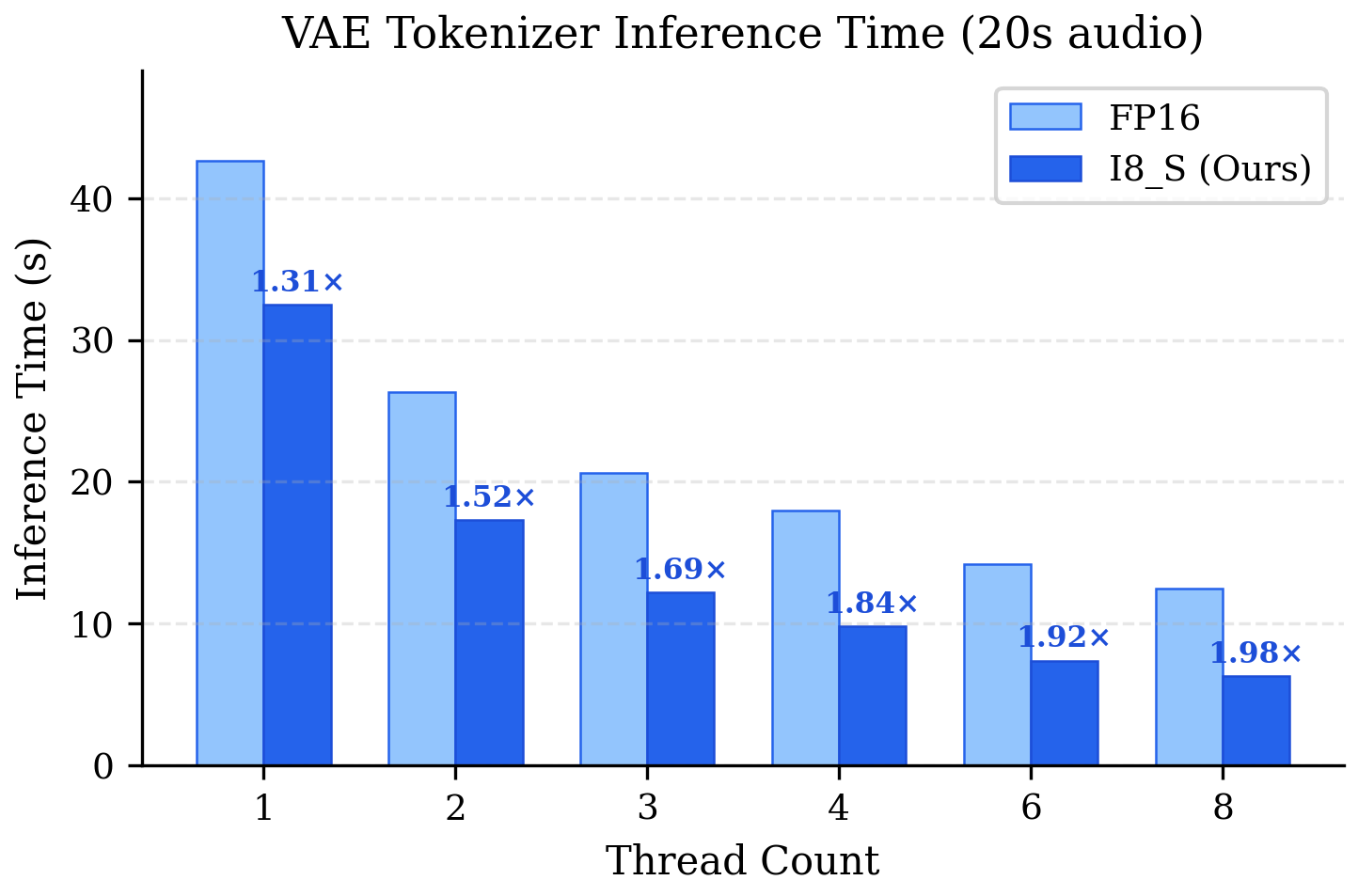}
\caption{VAE tokenizer inference time (s) vs.\ thread count (20\,s audio). Numbers above I8\_S bars indicate speedup over FP16.}
\label{fig:vae_speedup}
\end{figure}

To understand the source of speedup, we profile the VAE tokenizer and LM decoder separately. Figure~\ref{fig:vae_speedup} compares the VAE inference time under FP16 and I8\_S quantization. The I8\_S VAE achieves $1.3\times$--$2.0\times$ speedup, with the gain increasing at higher thread counts due to better SIMD utilization under reduced data width.

\begin{figure}[ht]
\centering
\includegraphics[width=\linewidth]{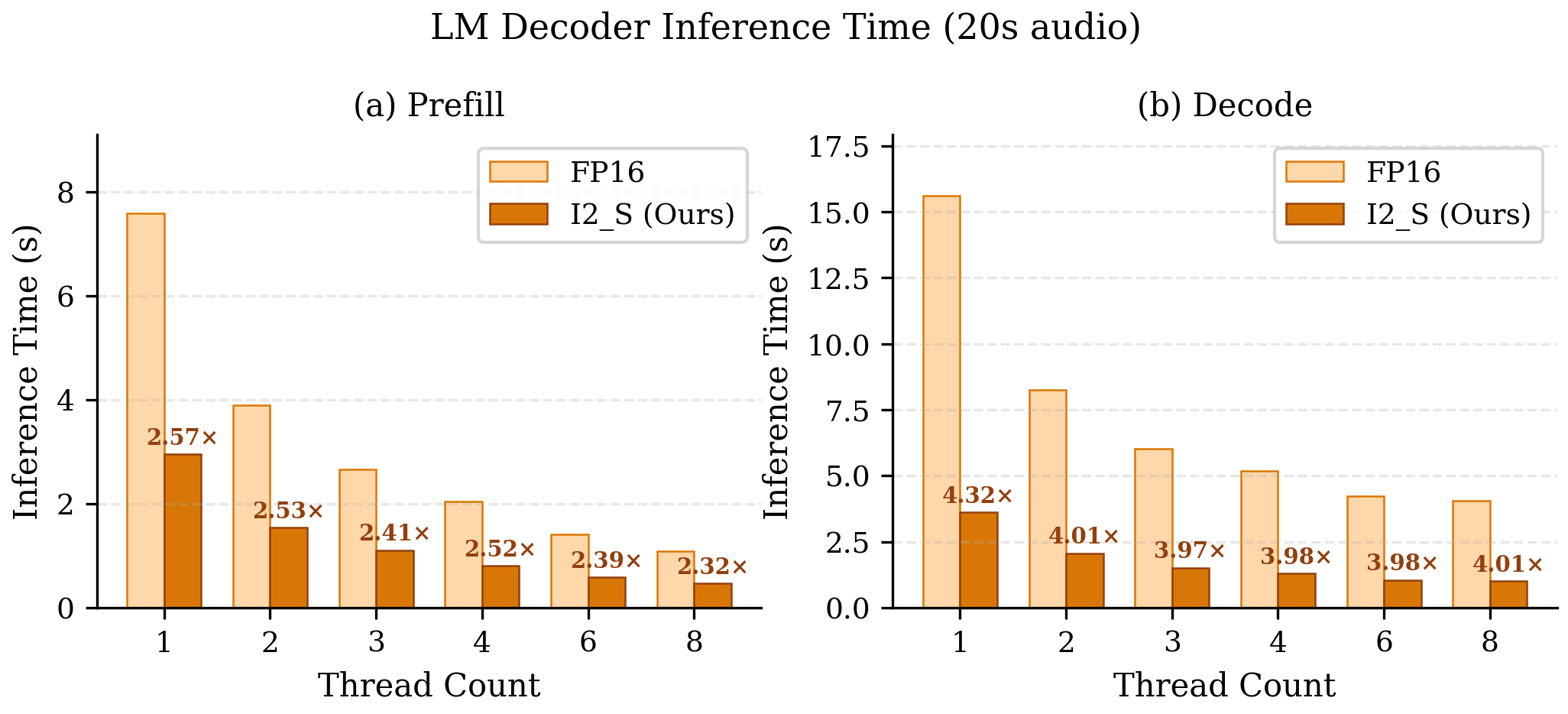}
\caption{LM decoder inference time (s) vs.\ thread count (20\,s audio). (a) Prefill stage. (b) Decode stage. Numbers above I2\_S bars indicate speedup over FP16.}
\label{fig:lm_speedup}
\end{figure}

Figure~\ref{fig:lm_speedup} presents the LM decoder profiling results. I2\_S achieves ${\sim}2.5\times$ speedup on prefill and ${\sim}4\times$ on decode. The higher decode speedup is expected: decode is purely weight-IO bound at batch size 1, where the $8\times$ weight compression of I2\_S relative to FP16 is most effective. FP16 decode scales poorly beyond 4 threads due to memory bandwidth saturation, while I2\_S continues to benefit from additional threads.

\subsection{Accuracy}
\label{sec:accuracy}

Table~\ref{tab:accuracy} compares WER and CER across 15 benchmarks. We include VibeVoice-ASR-7B as an upper-bound reference to quantify the accuracy gap introduced by model compression; however, since its 7B LM is $4.7\times$ larger than the other models in this comparison, it does not participate in the ranking. Among models of comparable size, \vibeasr{} achieves the lowest WER/CER on MLC-EN, AMI-ihm, AMI-sdm, and VoxPopuli. Despite the aggressive compression from 7B to 1.5B with ternary quantization, the accuracy degradation remains modest, typically 1--4\% absolute WER increase on most benchmarks. Parakeet leads on read-speech English benchmarks due to its English-specialized design, while SenseVoice and FunASR dominate Chinese benchmarks with Chinese-focused training data.

Notably, as an LM-based architecture, \vibebase{} natively supports multilingual recognition across 6+ languages---a capability absent in traditional ASR architectures such as Parakeet and FunASR-Nano, which are typically restricted to one or two languages. \vibeasr{} is also the only model in this comparison that runs entirely on CPU with heterogeneous quantization; all others require GPU inference for practical deployment.

\begin{table}[ht]
\centering
\renewcommand{\arraystretch}{1.15}
\newlength{\wercol}\settowidth{\wercol}{VibeVoice-}%
\resizebox{\textwidth}{!}{%
\begin{tabular}{cc|>{\centering\arraybackslash}p{\wercol}@{\hskip 4pt}>{\centering\arraybackslash}p{\wercol}|>{\centering\arraybackslash}p{\wercol}@{\hskip 4pt}>{\centering\arraybackslash}p{\wercol}|c@{\hskip 6pt}c|c@{\hskip 6pt}c|c@{\hskip 6pt}c|c@{\hskip 6pt}c}
\toprule
\multirow{2}{*}{} & \multirow{2}{*}{Benchmark} &
\multicolumn{2}{c|}{VibeVoice-ASR-7B} &
\multicolumn{2}{c|}{VibeVoice-ASR-BitNet$^*$} &
\multicolumn{2}{c|}{Parakeet} &
\multicolumn{2}{c|}{Whisper} &
\multicolumn{2}{c|}{SenseVoice} &
\multicolumn{2}{c}{FunASR} \\
& & WER & CER & WER & CER & WER & CER & WER & CER & WER & CER & WER & CER \\
\midrule
\multirow{6}{*}{\rotatebox{90}{MLC}}
& EN & 7.82 & 4.51 & \textbf{8.25} & \textbf{4.87} & 8.40 & 5.03 & 13.57 & 10.34 & 12.39 & 7.56 & 11.36 & 7.53 \\
& FR & 16.03 & 9.78 & 17.41 & 10.62 & --- & --- & --- & --- & --- & --- & --- & --- \\
& IT & 15.67 & 6.94 & 17.23 & 7.58 & --- & --- & --- & --- & --- & --- & --- & --- \\
& KO & 9.83 & 9.83 & 11.15 & 11.15 & --- & --- & --- & --- & --- & --- & --- & --- \\
& PT & 22.41 & 12.68 & 24.87 & 14.03 & --- & --- & --- & --- & --- & --- & --- & --- \\
& VI & 20.15 & 12.87 & 22.38 & 14.21 & --- & --- & --- & --- & --- & --- & --- & --- \\
\midrule
& AISHELL4 (ZH) & 19.83 & 19.83 & 27.45 & 27.45 & --- & --- & --- & --- & 22.52 & 22.52 & \textbf{20.41} & \textbf{20.41} \\
& AMI-ihm (EN) & 17.42 & 13.56 & \textbf{21.36} & \textbf{16.91} & 21.92 & 17.68 & 27.07 & 21.91 & 30.81 & 25.71 & 32.07 & 26.13 \\
& AMI-sdm (EN) & 24.18 & 18.73 & \textbf{25.87} & \textbf{20.94} & 26.33 & 21.26 & 36.92 & 31.37 & 48.11 & 42.67 & 40.17 & 32.39 \\
& AliMeeting (ZH) & 36.21 & 36.21 & 40.58 & 40.58 & --- & --- & --- & --- & \textbf{38.75} & \textbf{38.75} & 39.27 & 39.27 \\
& Fleurs-en (EN) & 4.73 & 2.05 & 5.21 & 2.28 & 4.09 & 1.67 & \textbf{3.99} & \textbf{1.64} & 6.84 & 2.81 & 4.93 & 2.20 \\
& Fleurs-zh (ZH) & 7.92 & 7.92 & 8.35 & 8.35 & --- & --- & --- & --- & \textbf{5.56} & \textbf{5.56} & 7.00 & 7.00 \\
& Libri-clean (EN) & 2.17 & 0.88 & 2.41 & 0.95 & \textbf{1.49} & \textbf{0.46} & 1.98 & 0.77 & 2.78 & 1.07 & 1.58 & 0.56 \\
& Libri-other (EN) & 5.84 & 2.94 & 6.27 & 3.18 & \textbf{3.13} & \textbf{1.16} & 3.60 & 1.52 & 6.81 & 3.31 & 4.01 & 1.77 \\
& VoxPopuli (EN) & 4.92 & 2.73 & \textbf{5.18} & \textbf{2.96} & 5.26 & 3.04 & 7.19 & 4.43 & 8.63 & 4.71 & 6.46 & 3.71 \\
\bottomrule
\end{tabular}%
}
\caption{WER/CER (\%) comparison. $^*$This work. Parakeet: Nvidia Parakeet-TDT-0.6B-v2 \citep{nvidia2024parakeet}; Whisper: OpenAI Whisper Large-v3 \citep{radford2023whisper}; SenseVoice: SenseVoice-small \citep{an2024sensevoice}; FunASR: FunASR-Nano \citep{gao2023funasr}. ``---'' indicates unsupported language or abnormally high error rate. Bold marks the best excluding VibeVoice-ASR-7B.}
\label{tab:accuracy}
\end{table}

\subsection{Comparison}
\label{sec:comparison}

To compare \vibeasr{}'s runtime efficiency with existing LM-based CPU ASR frameworks, we benchmark against Whisper.cpp on the same hardware using a 20-second English speech clip. Both systems share a similar architecture---encoder + autoregressive decoder---and comparable model sizes of $\sim$1.6\,GB, making this a direct comparison of quantization and runtime efficiency. Table~\ref{tab:speed} reports the results:

\begin{itemize}[nosep,leftmargin=1.5em]
    \item \textbf{\vibeasr{}}: VAE-I8\_S + LM-I2\_S, 1.6\,GB, 24\,kHz input.
    \item \textbf{Whisper.cpp large-v3-turbo}: INT8 quantization, $\sim$1.6\,GB, 16\,kHz input \citep{radford2023whisper}.
\end{itemize}

\begin{table}[ht]
\centering
\small
\begin{tabular}{c|ccccc}
\toprule
\textbf{Framework} & \textbf{1T} & \textbf{2T} & \textbf{4T} & \textbf{6T} & \textbf{8T} \\
\midrule
VibeVoice-ASR-BitNet$^*$ (1.6\,GB) & 44.8 & 25.2 & 15.3 & 11.5 & 10.0 \\
Whisper.cpp large-v3-turbo (1.6\,GB) & 102.1 & 53.3 & 28.4 & 19.7 & 15.5 \\
\midrule
Speedup & 2.28$\times$ & 2.12$\times$ & 1.86$\times$ & 1.71$\times$ & 1.55$\times$ \\
\bottomrule
\end{tabular}
\caption{End-to-end inference time (s) on 20\,s audio across thread counts. $^*$This work.}
\label{tab:speed}
\end{table}

\vibeasr{} is $1.6$--$2.3\times$ faster than Whisper.cpp across all thread configurations. The speedup is most pronounced at low thread counts, where the I2\_S ternary weight format maximally reduces memory bandwidth pressure. Both frameworks support multilingual recognition, but \vibeasr{} achieves this with a more aggressive quantization scheme---I8\_S + I2\_S vs.\ INT8-only---demonstrating that heterogeneous quantization can simultaneously improve both compression ratio and inference speed without sacrificing accuracy.

%% file: sections/conclusion.tex
\section{Conclusion}
\label{sec:conclusion}

We present VibeVoice-ASR-BitNet, a heterogeneous quantization framework that enables real-time LLM-based ASR inference on commodity CPUs. The core design applies full-pipeline INT8 quantization (I8\_S) to the activation-dominated VAE tokenizer, and BitNet ternary weights (I2\_S) to the weight-dominated LM decoder, matching each quantization scheme to the computational profile of its target component. A progressive QAT procedure with $\alpha$-blending ensures stable convergence under aggressive compression, while custom SIMD multiply-add kernels are tailored to both data formats. Together, these techniques compress the total model size from 4.62\,GB to 1.58\,GB ($2.9\times$ reduction), achieve real-time inference (RTF $< 1$) with modest thread budgets, and retain competitive multilingual accuracy across 6+ languages.

\vspace{0.5em}
\noindent\textbf{Limitations.}
(1)~The heterogeneous quantization strategy (I8\_S for VAE, I2\_S for LM) has only been validated on the VibeVoice-ASR architecture; its applicability to other VAE-LM or encoder-decoder ASR models (e.g., Whisper, Qwen-Audio) remains unexplored.
(2)~The current implementation supports offline (batch) inference only. Streaming mode---which requires chunked VAE encoding and incremental LM decoding---is not yet supported, limiting deployment in real-time interactive scenarios.